\documentclass{iaus}
\usepackage{graphicx}

\title[short title of paper] 
{Constraints on Bars in the Local Universe from 5000 SDSS Galaxies}

\author[short author list]   
{Fabio D. Barazza, Shardha Jogee, and Irina Marinova}

\affiliation{Department of Astronomy, University of Texas at Austin, 1
University Station C1400, Asutin, TX 78712-0259, USA \break
email: barazza@astro.as.utexas.edu}

\pubyear{2006}
\volume{235}  
\pagerange{119--126}
\date{?? and in revised form ??}
\jname{Galaxy Evolution across the Hubble Time}
\editors{F.Combes \& J. Palous, eds.}
\begin{document}

\maketitle

\begin{abstract}
We present the first study of bars in the local Universe, based on the Sloan
Digitized Sky Survey (SDSS). The large sample of $\sim5000$ local galaxies
provides the largest study to date of local bars and minimizes the effect of
cosmic variance. The sample galaxies have $M_g\leq -18.5$ mag and cover the
redshift range $0.01\leq z<0.04$. We use a color cut in the color-magnitude
diagram and the S\'ersic index $n$ to identify disk galaxies. We characterize
bars and disks using $r$-band images and the method of iterative ellipse fits
and quantitative criteria developed in Jogee at al. (2004, {\it ApJL}, 615,
L105). After excluding highly inclined ($i>60^{\circ}$) systems our results
are: (1) the optical ($r$-band) fraction of barred galaxies among local disk
galaxies is $43\%$ (Figure 1, left panel), which confirms the ubiquity of local
bars, in agreement with other optical studies based on smaller samples (e.g.
Eskridge et al. 2000, {\it AJ}, 119, 536, Marinova \& Jogee 2006,
astro-ph/0608039); (2) the optical bar fraction rises for bluer galaxies,
suggesting a relation between bars and star formation (Figure 1, middle panel);
(3) preliminary analyzes suggest that the optical bar fraction increases
steeply with the galaxy effective radius ($r_{eff}$, Figure 1, right panel);
(4) the optical bar fraction at $z\sim0$ is $\sim35\%$ for bright disks
($M_g\leq-19.3$ mag) and strong (bar ellipticity $>0.4$), large-scale (bar
semi-major axis $>1.5$ kpc) bars, which is comparable to the value of
$\sim30\pm6\%$ reported earlier (Jogee et al. 2004) for similar disks and bars
at $z\sim0.2-1.0$.
\begin{figure}[h]
\begin{center}
 \includegraphics{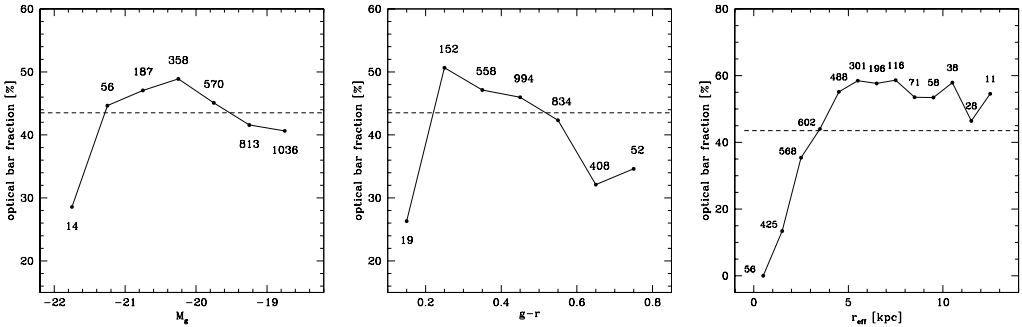}
  \caption{The optical bar fraction as a function of $M_g$ (left panel), $g-r$
color (middle panel), and $r_{eff}$ (right panel). The number next to each
point denotes the total number of galaxies in the corresponding bin. The dashed
line indicates the total optical bar fraction. Only bins with more than 10
objects are shown.}
\end{center}
\end{figure}
\end{abstract}

\firstsection 

\begin{acknowledgments}
We acknowledge support from NSF grant AST-0607748, NASA LTSA grant NAG5-13063,
and HST grant GO-10395.
\end{acknowledgments}

\end{document}